\documentclass{elsart4-1}

 \usepackage{graphicx}

\usepackage{amssymb}

\usepackage[english,francais]{babel}

\newtheorem{e-proposition}[theorem]{Proposition}

\newtheorem{e-definition}[theorem]{Definition\rm}

\renewcommand{\theequation}{\arabic{equation}}
\def\og{\leavevmode\raise.3ex\hbox{$\scriptscriptstyle\langle\!\langle$~}}
\def\fg{\leavevmode\raise.3ex\hbox{~$\!\scriptscriptstyle\,\rangle\!\rangle$}}

\begin{document}

\centerline{Physics or Astrophysics/Header}
\begin{frontmatter}


\selectlanguage{english}
\title{Speckle noise reduction techniques \\ for high-dynamic range imaging}


\selectlanguage{english}
\author[authorlabel1]{Pascal Bord\'e},
\ead{pascal.borde@ias.u-psud.fr}
\author[authorlabel2]{Wesley Traub}
\ead{wesley.a.traub@jpl.nasa.gov}

\address[authorlabel1]{Institut d'astrophysique spatiale (IAS), b\^atiment 121, Universit\'e Paris-Sud 11 and CNRS (UMR 8617), 91405 Orsay, France}
\address[authorlabel2]{Jet Propulsion Laboratory (JPL), MS 301-451, 4800 Oak Grove Drive, Pasadena, CA 91109, USA}


\medskip
\begin{center}
{\small Received *****; accepted after revision +++++}
\end{center}

\begin{abstract}
High-dynamic range imaging from space in the visible, aiming in particular at the detection of terrestrial exoplanets, necessitates not only the use of a coronagraph, but also of adaptive optics to correct optical defects in real time. Indeed, these defects scatter light and give birth to speckles in the image plane. Speckles can be cancelled by driving a deformable mirror to measure and compensate wavefront aberrations. In a first approach, \emph{targeted speckle nulling}, speckles are cancelled iteratively by starting with the brightest ones. This first method has demonstrated a contrast better than $10^9$ in laboratory. In a second appraoch, \emph{zonal speckle nulling}, the total energy of speckles is minimized in a given zone of the image plane. This second method has the advantage to tackle simultaneously all speckles from the targeted zone, but it still needs better experimental demonstration.
{\it To cite this article: P. Bord\'e, W. Traub, C. R. Physique 6 (2005).}

\vskip 0.5\baselineskip

\selectlanguage{francais}
\noindent{\bf R\'esum\'e}
\vskip 0.5\baselineskip
\noindent
{\bf Techniques de r\'eduction du bruit de tavelures pour l'imagerie \`a haute dynamique. }
L'imagerie \`a haute dynamique dans le visible et depuis l'espace, visant en particulier \`a la d\'etection d'exoplan\`etes telluriques, n\'ecessite non seulement l'utilisation d'un coronographe, mais aussi d'une optique adaptative pour corriger les d\'efauts optiques en temps r\'eel. En effet, ces d\'efauts diffusent la lumi\`ere et donnent naissance \`a des tavelures dans le plan image. Les tavelures peuvent \^etre supprim\'ees en pilotant un miroir d\'eformable pour mesurer, puis compenser les aberrations du front d'onde. Dans une premi\`ere approche, \emph{la suppression cibl\'ee de tavelures}, les tavelures sont supprim\'ees it\'erativement en commen\c{c}ant par les plus brillantes. Cette premi\`ere m\'ethode a permis d'atteindre un contraste de $10^9$ en laboratoire. Dans une deuxi\`eme approche, \emph{la suppression zonale de tavelures}, l'\'energie totale du champ de tavelures est minimis\'ee dans une certaine zone du plan image. Cette seconde m\'ethode a pour avantage de s'attaquer simultan\'ement \`a toutes les tavelures de la zone vis\'ee, mais elle doit encore faire ses preuves exp\'erimentalement.
{\it Pour citer cet article~: P. Bord\'e, W. Traub, C. R. Physique 6 (2005).}

 \vskip 0.5\baselineskip
\noindent{\small{\it Mots-cl\'es~:} Exoplan\`etes~; Coronographie~;
Tavelures}}
\end{abstract}
\end{frontmatter}

\selectlanguage{francais}
\section*{Version fran\c{c}aise abr\'eg\'ee}

\selectlanguage{english}

\section{Introduction} \label{sec:introduction}
In any optical instrument, optical aberrations (e.g., polishing defects or coating defects) scatter light in the image plane. This redistributed light -- known as \emph{speckle noise} -- reduces both the sensitivity and the angular resolution of the instrument. Even more light is scattered in the image plane when the observation takes place through a turbulent medium like the atmosphere. In order to recover the lost sensitivity and angular resolution, astronomers have developed active and adaptive optical systems \cite{Roddier99}. These systems make use of deformable mirrors (DMs) driven to compensate optical aberrations as they come and go. Active optics refers to slowly varying aberrations (e.g., thermal or gravity induced) and adaptive optics to quickly varying aberrations (e.g., atmospheric).

For the past decade, the quest for exoplanet imaging has been a strong driver to improve the sensitivity and angular resolution of astronomical instruments. Exoplanet imaging involves besides a dynamical range problem due to the angular proximity of the exoplanet parent star which is much brighter than the exoplanet itself (typically by a factor $10^{10}$ in the visible). Numerous optical solutions -- collectively known as \emph{coronagraphs} -- have been devised to overcome this dynamical range problem by blocking the light of the star while letting through the light of the exoplanet \cite{Guyon06}. Coronagraphs draw their names from the original instrument that Bernard Lyot built to observe the faint solar corona by artificially eclipsing the Sun \cite{Lyot32}. The most exciting and challenging application of coronagraphic telescopes is arguably the imaging and spectroscopy of terrestrial-size exoplanets, which seems -- to date -- more feasible from space than from the ground. NASA's Terrestrial Planet Finder Coronagraph \cite{Traub06} is one of the most advanced projects with that goal in the visible range of the spectrum.

In this paper, we discuss strategies to combine coronagraphs and wavefront control systems in order to achieve both the sensitivity and angular resolution required by terrestrial-size exoplanet imaging from space. Information about ground-based instruments can be found elsewhere \cite{Dohlen06}. After a few basic ideas about coronagraphs (\S\ref{sec:coronagraphs}) and the description of the mechanism of speckle formation (\S\ref{sec:speckle_formation}), we present two schemes for speckle noise reduction, a.k.a. speckle nulling : targeted speckle nulling (\S\ref{sec:targeted_sn}) and zonal speckle nulling (\S\ref{sec:zonal_sn}).

\section{Basic ideas about coronagraphs} \label{sec:coronagraphs}
To date, most of the work pertaining to speckle nulling techniques for coronagraphs has been achieved for either Lyot-type coronagraphs or shaped-pupil coronagraphs. Here, we briefly describe these two types.

\subsection{Lyot-type coronagraphs}
\begin{figure}
\centering
\includegraphics[width=12cm]{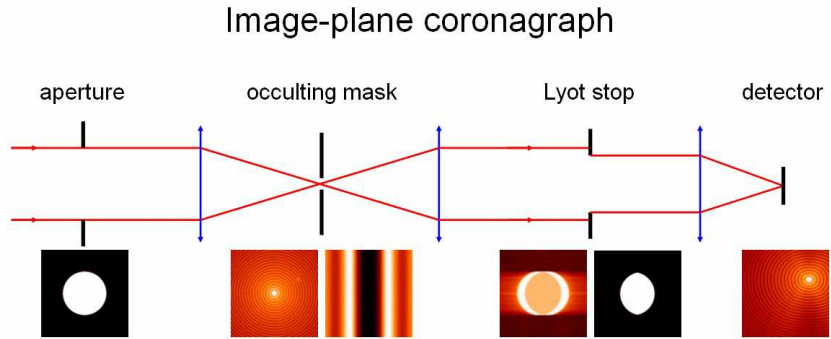}
\caption{Image-plane coronagraph with band-limited amplitude mask}
\label{fig:fig1}
\end{figure}

Lyot-type coronagraphs consist in two-stage systems comprising an image-plane mask followed by a pupil-plane diaphragm (Fig.~\ref{fig:fig1}). In the traditional design, the mask is a few Airy ring wide opaque disk that blocks the light from on-axis sources. Most of the light diffracted by the edge of the mask is then prevented from reaching the final focus by a diaphragm. In recent concepts, image-plane masks act either on the phase or the amplitude of on-axis light so as to diffract it entirely where the diaphragm filters it out of the system. For instance, four-quadrant phase mask coronagraphs \cite{Rouan00} and band-limited amplitude mask coronagraphs \cite{Kuchner02} can be shown to extinct perfectly on-axis light (in the approximation of Fourier optics).

\subsection{Shaped-pupil coronagraphs}
Shaped-pupil coronagraphs \cite{Kasdin03} consist also in two-stage systems where, this time, incoming light encounters first a pupil-plane diaphragm (or shaped pupil) followed by an image-plane mask. Shaped pupils are designed so as to diffract almost no light in certain directions, i.e., their diffraction patterns have very dark zones, contrary to circular apertures that diffract light in all directions equally. Matching image-plane masks are then placed right in front of the detector to prevent imaging of the strongly illuminated regions.

\subsection{The need for wavefront correction}
Thus coronagraphs reduce dramatically the level of diffracted light of an on-axis stellar source and its associated photon noise. However, as we mentioned in \S\ref{sec:introduction}, optical aberrations bring back stellar light in the final focus and give rise to speckle noise. For this reason, coronagraphs must be associated to wavefront correction devices, such as DMs. This is mandatory -- even in space -- to achieve the $10^{10}$ necessary contrast for the successful imaging of terrestrial-size exoplanets. Note that wavefront analysis needs to be done with the science channel, since a separate analysis channel would introduce its own aberrations, difficult to disentangle from aberrations of the science channel.

\section{Speckle formation}  \label{sec:speckle_formation}
\begin{figure}
\centering
\includegraphics[width=12cm]{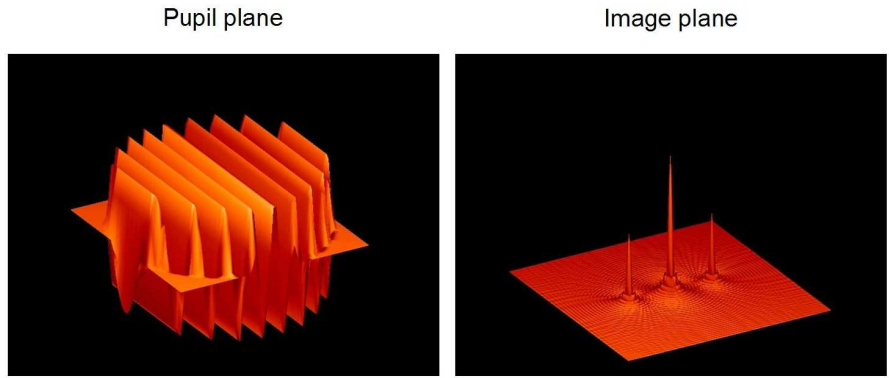}
\caption{Principle of speckle formation: phase aberrations can be expanded into Fourier components; each of these components acts as a phase grating that produces two symmetric speckles in the image plane.}
\label{fig:fig2}
\end{figure}

As a starting point, let us consider an optical surface with uniform reflectivity but with polishing errors resulting in deviations (e.g., holes and bumps) from the desired surface. The bidimensional map of these deviations can be expanded into Fourier components. Any of these components is like a physical ripple on the surface that acts as a phase grating and diffracts light in specific directions that depend on the orientation and spatial frequency of the component considered. In the image plane, every order of diffraction of each of these phase gratings appear as a spot that we call speckle (Fig.~\ref{fig:fig2}).

Heterogeneities in the reflectivity of optical surfaces can be expanded in the same manner, and act as amplitude gratings with associated speckles in the image plane. One can show that propagation mix phase and amplitude errors \cite{Guyon05}, so that both types are generally present in the final focus (although amplitude errors are usually smaller). 

\section{Targeted speckle nulling}  \label{sec:targeted_sn}
\subsection{Principle}
From \S\ref{sec:speckle_formation}, we conclude that speckles would disappear if, for every Fourier component, we could impose on an active element, e.g., a DM for phase correction only and two DMs in a Michelson interferometer configuration for phase and amplitude correction, the same Fourier component with opposite amplitude. For this to work, we need to extract from the image the amplitude and phase of each Fourier component. Neglecting orders of diffraction beyond $\pm 1$ (i.e., assuming small aberrations), each Fourier component appears in the image plane as a pair of speckles symmetric with respect to the optical axis. In the case of phase aberrations only, both speckles have an intensity equal to half the square of the amplitude of the Fourier component. The phase of the Fourier component, however, cannot be retrieved from the intensity map of the image plane. Therefore, this phase must be found by trial and error. It is easy to show that enough information is acquired after two trials to be able to compute the right phase, so that the Fourier component should disappear from the fourth image. When there are also amplitude aberrations, the intensity of the two speckles of the pair differ, but a series of three images still contains enough information to fully characterize the Fourier components.

In practice, not all Fourier components can be adressed at once, so this method works iteratively by targeting the brightest speckles at each measurement and correction cycle. This is the reason why we refer to this method originally developed by Chris Burrows as \emph{targeted speckle nulling}.

\subsection{Laboratory experiments}
Targeted speckle nulling has been in use for several years on the High Contrast Imaging Testbed (HCIT) at the Jet Propulsion Laboratory, and has made it possible to reach repeatedly a contrast better than $10^9$ in monochromatic light with a band-limited coronagraph \cite{Trauger04,Trauger06}. More research is underway to attain such a high contrast level with this technique in polychromatic light. A first experiment in monochromatic light with a shaped-pupil coronagraph has also been conducted on the HCIT by \cite{Belikov06a}, and demonstrated a contrast of $2.5 \!\times\! 10^7$.

\section{Zonal speckle nulling}  \label{sec:zonal_sn}
\subsection{Principle}
Zonal speckle nulling \cite{Borde06a,Borde06b} has been developed as an attempt to tackle all speckles at every measurement and correction cycle, in order to reduce as much as possible the number of cycles needed to reach a given contrast. It builds on the \emph{dark-hole algorithm} proposed by \cite{Malbet95}, but includes an original measurement strategy. The fundamental difference with targeted speckle nulling described above is that, instead of measuring amplitudes and phases of individual speckles, the algorithm solves for the amplitude and phase of the electric field in every \emph{detector pixel}. Once the map of the electric field is known, a couple of numerical methods can be used to \emph{compute the DM shape that minimizes the total energy of speckles inside a given zone}.

\begin{figure}
\centering
\includegraphics[width=6.5cm]{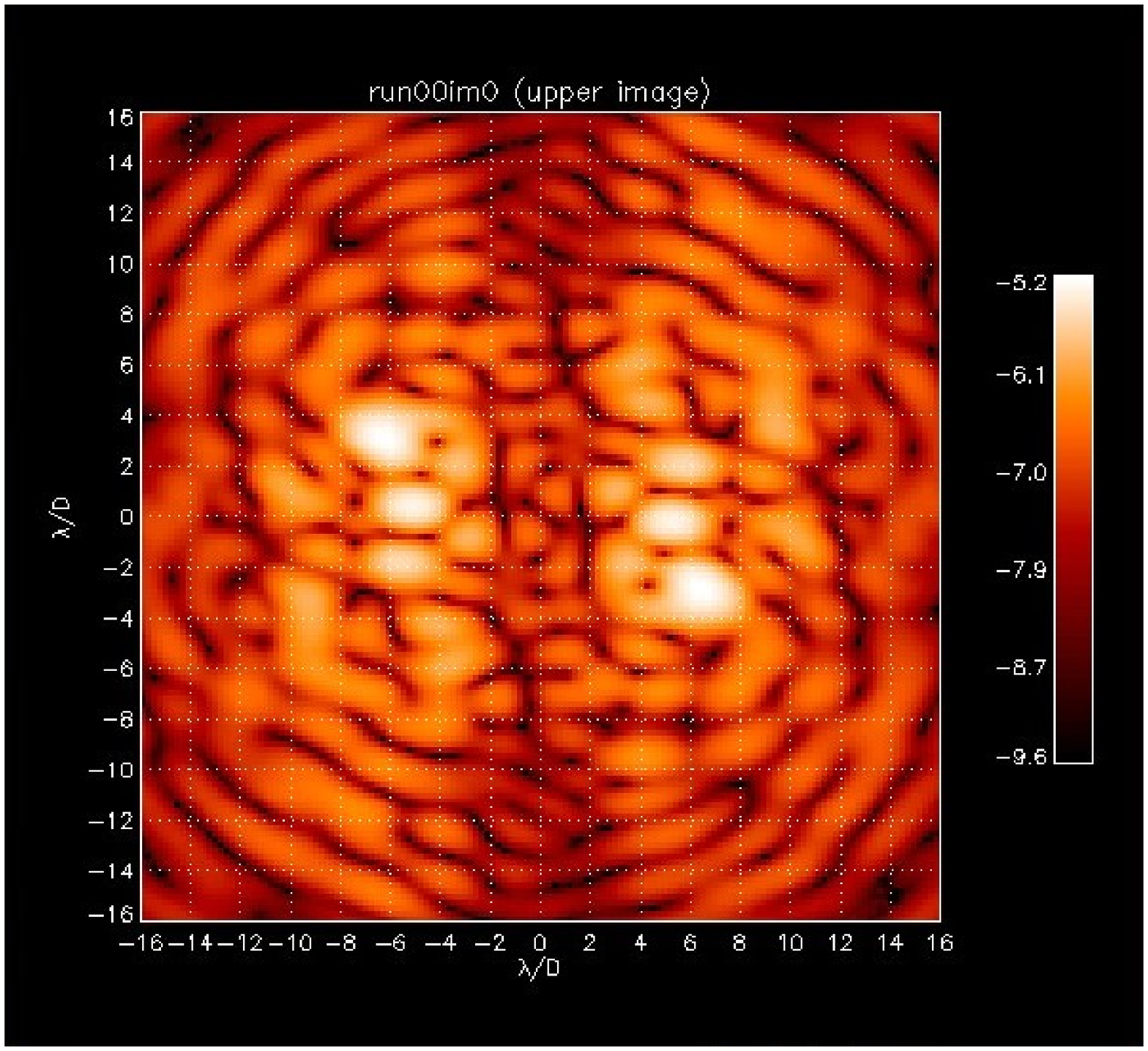}
\includegraphics[width=6.5cm]{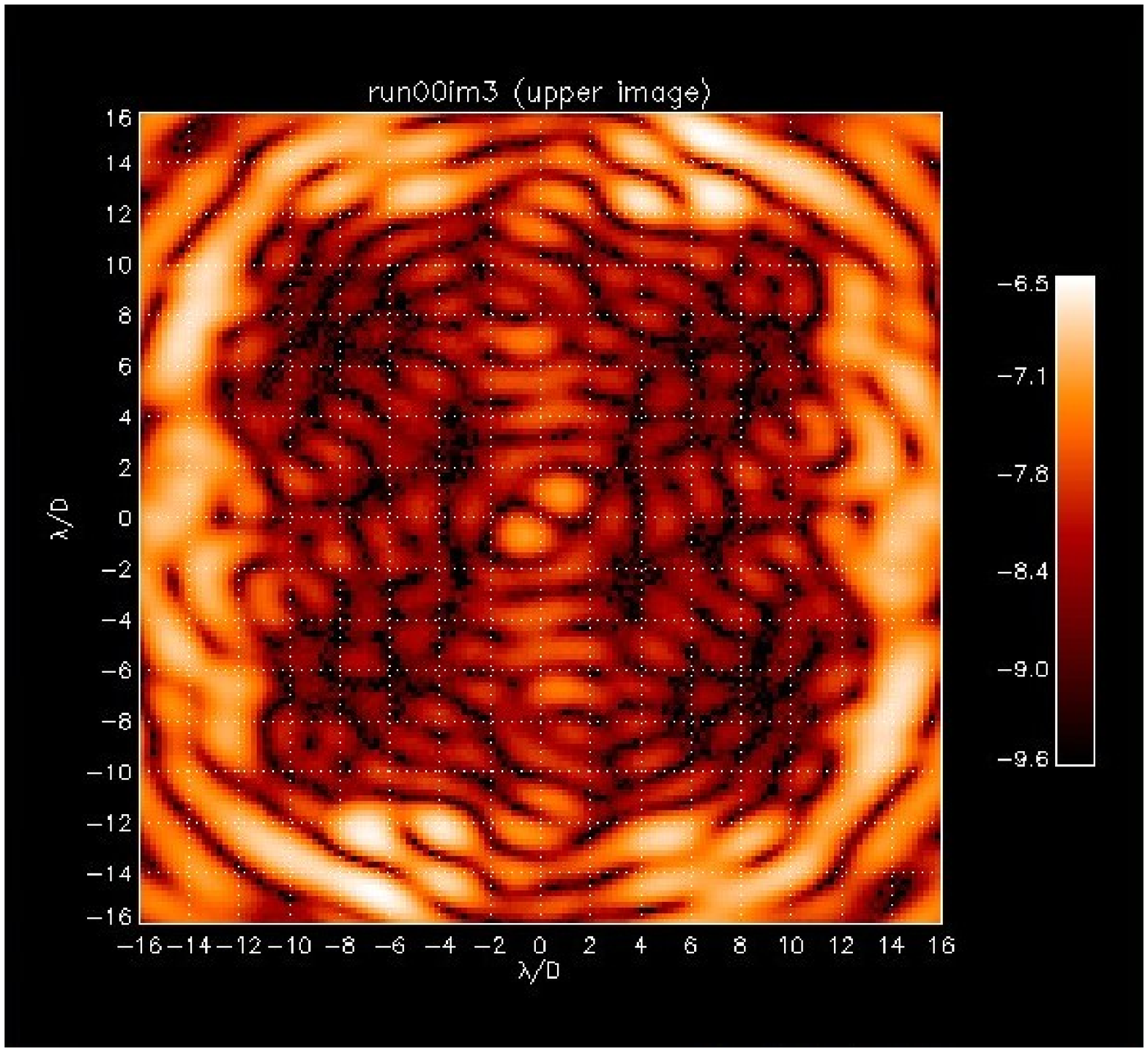}
\includegraphics[width=6.5cm]{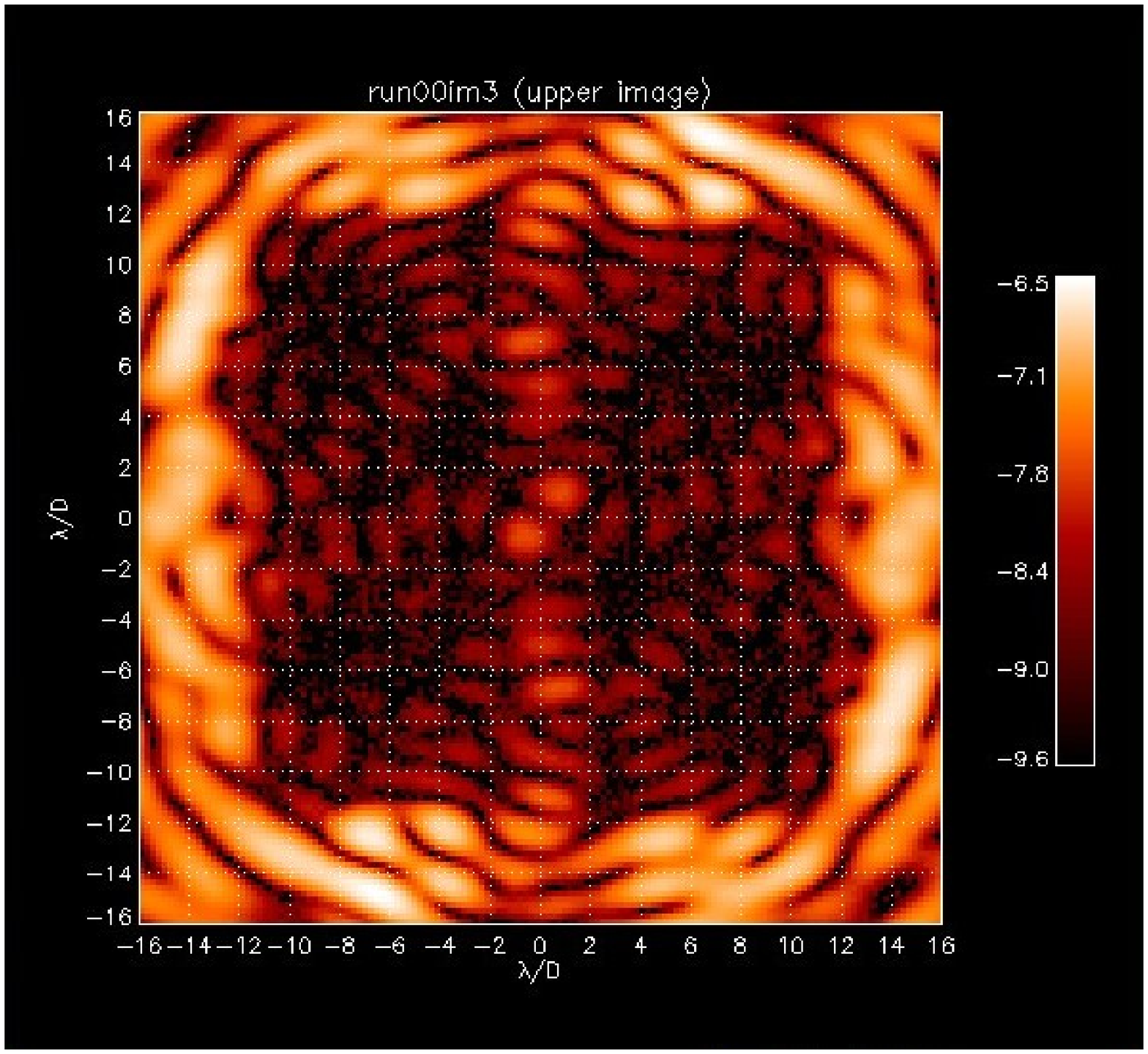}
\includegraphics[width=6.5cm]{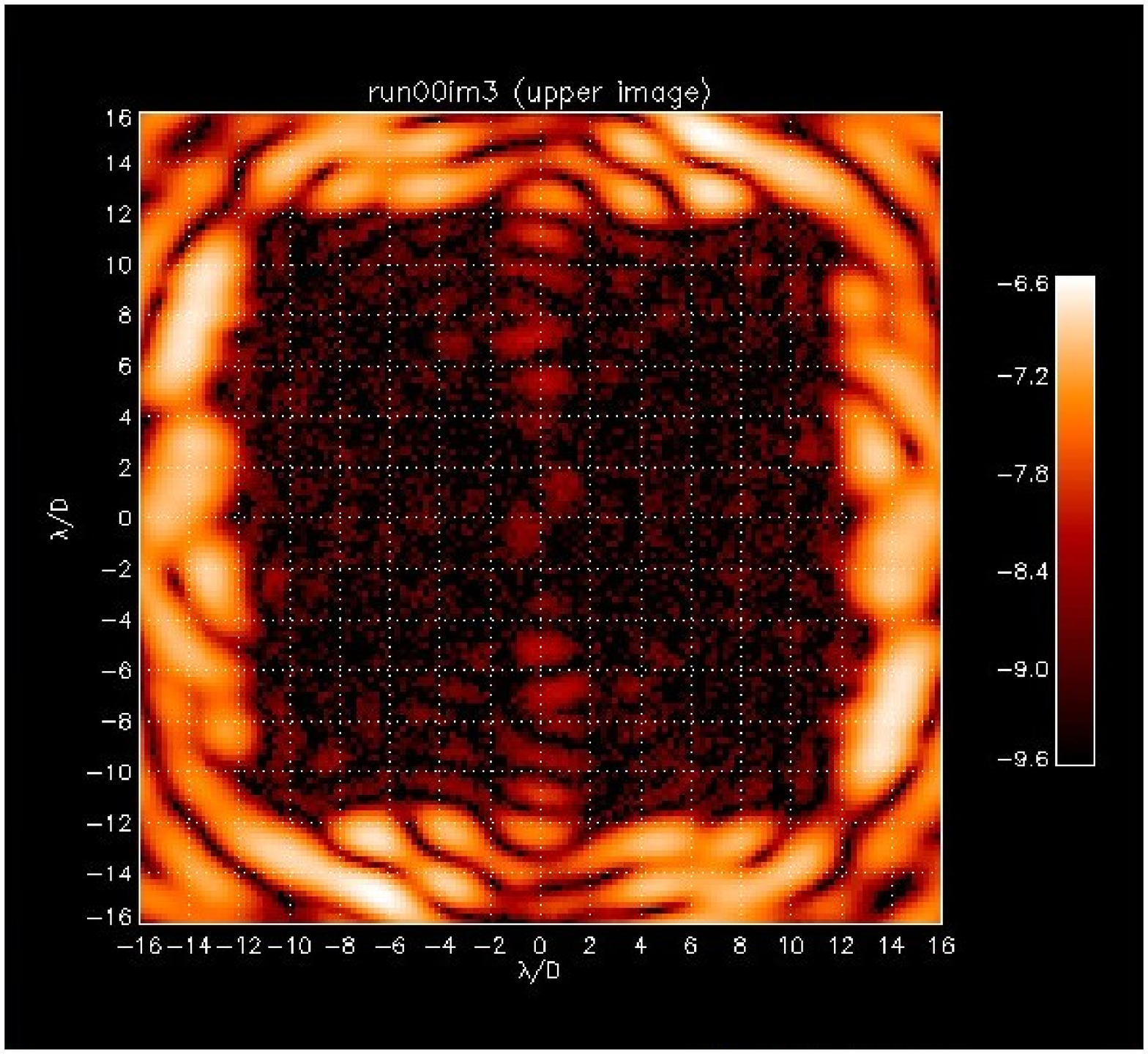}
\includegraphics[width=6.5cm]{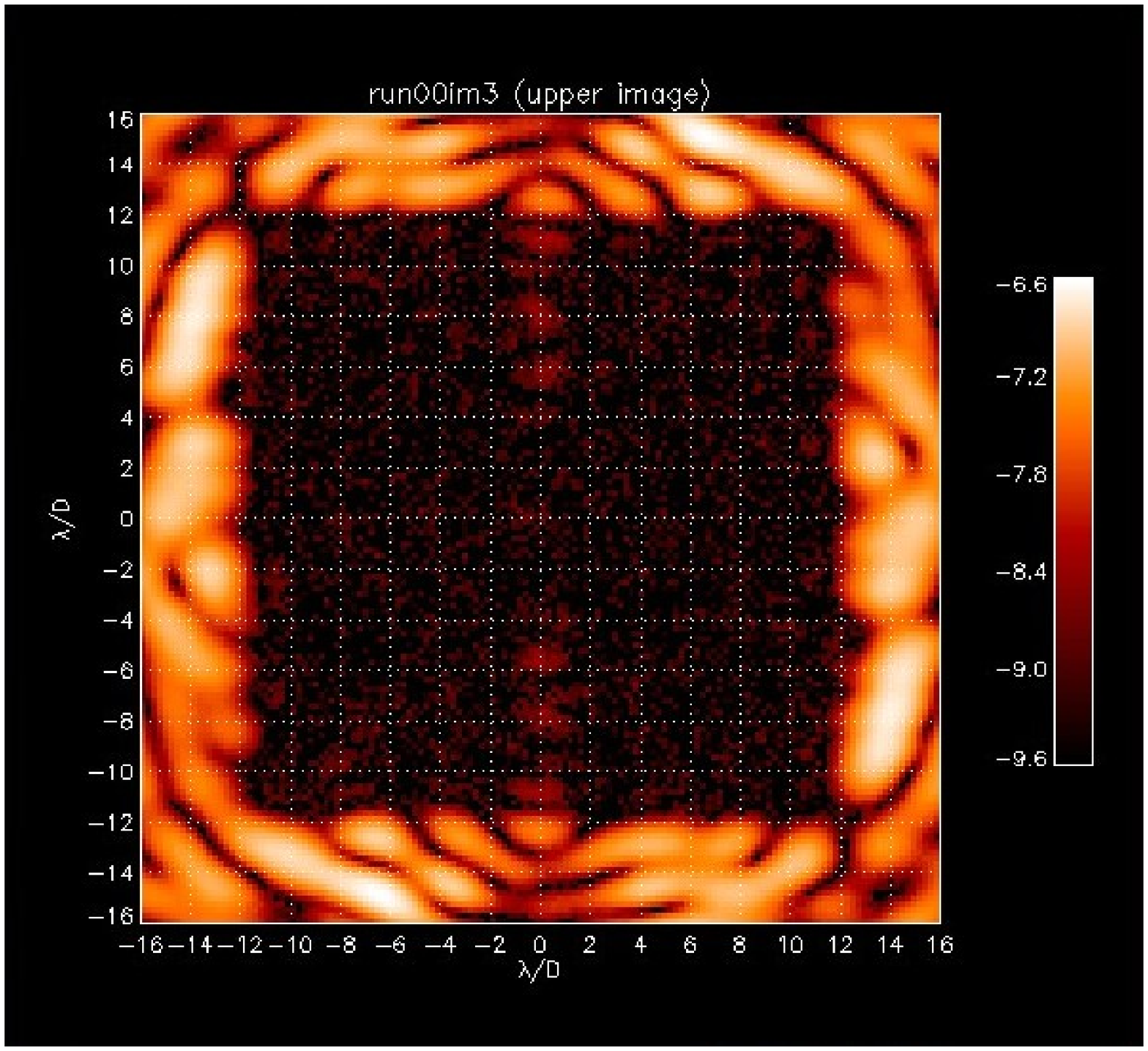}
\includegraphics[width=6.5cm]{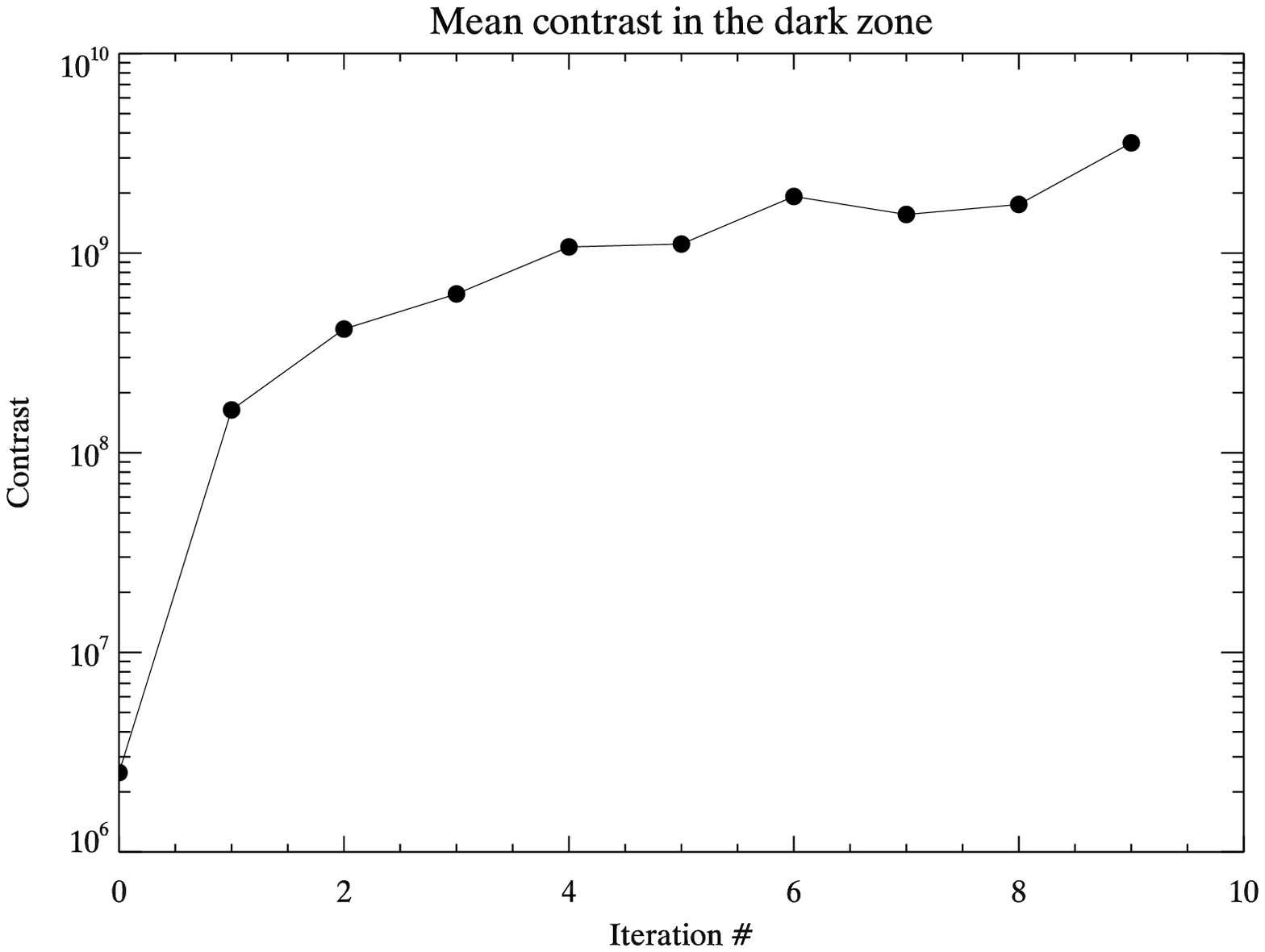}
\caption{Simulation of zonal speckle nulling with the HCIT and a band-limited coronagraph of the type in Fig.~\ref{fig:fig1}. We consider phase aberrations only with a power spectral density decreasing as the third power of the spatial frequency. The deformable mirror has $32 \!\times\! 32$ actuators placed on a square grid. Although speckles could be nulled in a zone $32 \!\times\! 32$ elements of resolution ($\lambda/D$) wide with an ideal coronagraph, the targeted zone is here a rectangle from 1 to 12~$\lambda/D$ in width, and from $-12$ to 12~$\lambda/D$ in height. Indeed, speckles are more difficult to correct on a vertical strip where the mask transmission is low, and at spatial frequencies approaching the Nyquist frequency of the DM. Images are color-coded on a log scale featuring the inverse of the contrast (i.e., $-10$ corresponds to a contrast of $10^{10}$). The first image shows the uncorrected speckle field. The second, third, fourth, and fifth images show the corrected field after one, two, four, and ten iteration(s), respectively. The plot in the bottom-right corner displays the mean contrast in the dark zone as a function of the number of iterations. In this simulation, the contrast in a given pixel is computed by dividing the intensity in that pixel by the peak intensity of the star observed through the Lyot stop without the mask. The contrast is not normalized by the mask transmission that amounts to about 50~\% at 4~$\lambda/D$ and 80~\% at 5.2~$\lambda/D$.}
\label{fig:fig3}
\end{figure}

Because the algorithm is built on the hypothesis of small aberrations, perfect speckle cancellation cannot be obtained in one measurement and correction cycle, but the algorithm could be iterated as demonstrated in Fig.~\ref{fig:fig3}. Moreover, because the algorithm relies on models of the DM and the coronagraph to compute the effect -- in the final image plane -- of the actuation of a given DM actuator, these models should be made as realistic as possible for an accurate speckle nulling. For instance, they should include the detailed effect of the real coronagraph on the phase and the amplitude of the electric field, or subtle phenomena such as electronic cross-talk between actuactor drivers. Any unmodeled effect could be treated as additional phase and/or amplitude aberrations, but will necessitate more iterations.

Like for any other methods, the zone where speckle are nulled is limited in extension by the spatial resolution of the DM. Precisely, a square DM with $N\!\times\!N$ actuators makes it possible to suppress speckles in a square zone of the image plane that is $N$ elements of resolution wide, provided only phase aberrations are present. If there are also amplitude aberrations and a single DM is used, then only half of the zone can be corrected. However, the full zone could be corrected either by placing two DMs in a Michelson interferometer configuration, or by inserting in the optical train a second DM at a location intermediate between a pupil and an image plane and taking advantage of Fresnel propagation effects \cite{Shaklan06}.

Zonal speckle nulling has also been adapted to shaped-pupil coronagraphs with a measurement strategy using a pinhole instead of the DM as the tool for phase diversity \cite{Giveon06}.

\subsection{Laboratory experiments}
Preliminary zonal speckle nulling experiments have been conducted on the HCIT with a band-limited coronagraph in monochromatic light, and encouraging results have been obtained \cite{Borde06c}. A testbed featuring a shaped-pupil coronagraph and a DM is being developed at Princeton University \cite{Belikov06b}, and more experiments will be conducted on the HCIT with shaped-pupils in the near future. 

\section{Conclusion} \label{sec:conclusion}
Space-born coronagraphic telescopes need to be equipped with adaptive optics in order to attain contrast ratios compatible with terrestrial exoplanet detection. Indeed, coronagraphs decrease the photon noise due to diffracted star light, while adaptive optics can be driven to decrease speckle noise due to star light scattered by optical train defects. We have presented two approaches to suppress speckles behind coronagraphs: targeted speckle nulling and zonal speckle nulling. On the one hand, targeted speckle nulling benefit from several years of development and experiments, but it necessitates a number of iterations that might make it unpractical. On the other hand, zonal speckle nulling should necessitate fewer iterations, but still lacks a clear experimental demonstration.

We note that different approaches for speckle nulling have been proposed, like \emph{synchronous interferometric speckle subtraction} \cite{Guyon04}, or \emph{anti-halo apodization} \cite{Codona04}. In these approaches, speckles are suppressed by diverting some of the star light before it enters the coronagraph, and using it to construct an anti-speckle field that is added to the image plane. Experimental demonstration is also underway \cite{Putnam06}.




\end{document}